# Evolution Of Entanglement In Groverian Search Algorithm: Two-Qutrit System

Arti Chamoli and C. M. Bhandari

Indian Institute of Information Technology, Allahabad, Uttar Pradesh, India

Entanglement plays a crucial role in quantum processes particularly those pertaining to quantum information and computation. An analytic expression for entanglement measure defined in terms of success rate of Grover's search algorithm has been obtained for a two-qutrit system and the calculated results agree well with the conventional entropy based measure. Qutrit systems are of special interest because the Hilbert space dimensionality (for a given number of qudits- d-dimensional system) is optimal for d=3 and this may be of significance in the enhancement of computing power.

## PACS

03.65.Ud, 03.67.–a,

## Introduction

Quantum entanglement is the heart and soul of quantum information processing. Its role is distinct and explicit in some situations such as teleportation and superdense coding, whereas it is not so explicit in search algorithms. For obvious reasons most of research effort so far has been focussed on quantum bits or qubits. The simplicityof two-state systems and the ease with which they can be handled has been primarily responsible for this. However, in principle there is no reason to limit quantum information and computation architectures to two-level systems. The total dimensionality of Hilbert space can be increased by considering qudits; a d-level quantum system which takes d=2 for qubits. The next reasonable step is to look for quantum computation architectures with d=3 which is a qutrit or a three level qudit. During recent years several researchers have investigated such systems in different contexts. Generation and characterization of entanglement for three level system[1]; quantum key distribution protocol with qutrits[2]; quantum tomography for qudits[3]; entanglement swapping between multi qudits[4]; discrimination among Bell states of qudits[5]; GHZ paradox for many qudits[6]; quantum computing with qudits[7]; quantum communication complexity protocol with two entangled qutrits[8]; bounds of entanglement between qudits[9]; entanglement among qudits[10] are worth mentioning in this context. All this unambiguously marks entanglement as the marrow of theory of information and computation. Thus quantification of the same gains utmost importance. Consequently many entanglement measures have been proposed for qubits to date [11-18]. Recently, F Pan et al [19] presented a classification for entangled bipartite qutrit states based on an entanglement measure[20]. In [21] authors have obtained an entanglement measure for certain kind of four qubit states. Entanglement in a real four qubit system was expressed in terms of its success probability as initial state of Grover's search algorithm. The measure meets the requirements of being zero for a product state and being invariant under local unitary transformations. Following the same line of thought a measure of entanglement has been framed for a two qutrit system. A qutrit is a unit of quantum information whose substates exist in a three dimensional Hilbert space. Accordingly, an n-qutrit system can have $3^n$ subsystems parallely. Violation of local realism being escalated in qutrit correlations enhances its suitability for tasks like cryptography. Use of qutrit systems makes the quantum cryptography protocols robust against eavesdropping attack[2,22-23]. In addition, security of quantum bit commitment and coin-flipping protocols is higher with entangled qutrits[24]. All these breakthroughs have motivated the development of an entanglement measure for d-dimensional systems, where d > 2. A qutrit system is of special interest because information processing appears to have great potential in a three-level system as it best fits into dimensionality aspect of Hilbert space[25]. The Hilbert-space dimensionality is maximized for d=3, and hence the computing power.

**Grover's search algorithm and entanglement measure**

We start with a brief review of modified unsorted database search[26], thereby expressing entanglement measure as a derivative of search algorithm. Search space for n qubits has $N$ elements such that $N = 2^n$. Thus the elements can be represented by an n-qubit register. Out of these $N$ elements a subset of r elements is marked and we wish to search the whole space in order to find a marked element. The state of n qubits is given by the state $|\phi\rangle$. The search algorithm proceeds with the introduction of an ancilla qubit $|0\rangle_q$ along with the input register $|\phi\rangle$, in the following way:

1. A product of arbitrary local operations, $V = U_1 \otimes U_2 \otimes --- U_n$ on the register and the gate HX on the the ancilla qubit is applied.

$$V |\phi\rangle \otimes HX |0\rangle_q \qquad (1)$$

2. Now the marked state is rotated by a phase of $\pi$ radians. Next all register states are rotated by $\pi$ radians around the average amplitude of the register state.

These two operations constitute Grover iteration $U_G$. The Grover iterations are applied m times, till the amplitude of the marked state reaches a maximum value.

3. Finally, the register is measured in the computational basis. If $P_{max}$ be the maximal success probability of search algorithm where maximization is over all possible local unitary operations in the initial step, then $P_{max}$ can be written in terms of $U_G^M$ (i.e. m grover iterations). $P_{max}$ is then obtained by averaging uniformly over all $N$ possible value for s($|s\rangle$ being the marked state).

Thus,

$$P_{max} = \max_{U_1 ----- U_n} \frac{1}{N} \sum_{s=0}^{N-1} \left| \langle s | U_G^m \left( U_1 \otimes U_2 \otimes ---- \otimes U_n \right) | \phi \rangle \right|^2 \qquad (2)$$

For a general state, we consider the effect of the Grover iterations on an uniform superposition state $|\eta\rangle = \sum_x \frac{|x\rangle}{\sqrt{N}}$. Applying m Grover iterations to this state yields

$$U_G^m |\eta\rangle = |s\rangle + O\left(\frac{1}{\sqrt{N}}\right) \qquad (3)$$

Second term is a small correction term because the solution of iterrative process is probabilistic rather than deterministic.

Multiplying by $\left(U_G^m\right)^\dagger$ and taking the hermitian conjugate gives

$$\langle s | U_G^m = \langle \eta | + O\left(\frac{1}{\sqrt{N}}\right) \qquad (4)$$

Substituting eq.(4) in eq.(2) gives

$$P_{max} = \max_{U_1 - - - - U_n} \frac{1}{N} \sum_{s=0}^{N-1} \left| \langle \eta | U_1 \otimes U_2 \otimes - - - - \otimes U_n | \phi \rangle \right|^2 + O\left( \frac{1}{\sqrt{N}} \right)$$

(5)

$|\eta\rangle$ being a product state implies that $U_1^\dagger \otimes U_2^\dagger \otimes - - - - \otimes U_n^\dagger |\eta\rangle$ is also a product state.

Thus optimization can be considered over product states. Hence,

$$P_{max} = \max_{|e_1 - - - - e_n\rangle} \left| \langle e_1 - - - - e_n | \phi \rangle \right|^2 + O\left( \frac{1}{\sqrt{N}} \right)$$

(6),

$|e_1\rangle \otimes - - - |e_n\rangle$ being a product state of n qubits.

If the input state were a product state, only then $P_{max}$ would be one upto some small corrections. This suggest that $P_{max}$ is affected by the entanglement of the initial state $|\phi\rangle$

Hence based on maximum success probability $P_{max}$, an entanglement measure was framed in [26,27]. This was followed by authors in [21] to study entanglement for a four qubit system. In the present letter, the same approach has been extended to three level system. An analytical expression giving entanglement measure for a two qutrit system is derived from modified Grover's search algorithm.

According to the definition [26], the entanglement measure $G(\psi)$ in terms of maximum success probability $P_{max}$ can be expressed as

$$G(\psi) = \sqrt{1 - P_{max}}$$

(7)

Dependence on $P_{max}$ makes it obvious that $G(\psi)$ will have its values in the range $0 \leq G(\psi) \leq 1$.

**Groverian entanglement measure for qutrit systems**

A d-level system represents a qudit. Thus there are d states which can be labeled as $|k\rangle$ $(k = 1, 2 - - - d)$. A general state of a qudit in a d-dimensional Hilbert space $H_d$ can thus be written as

$$|\xi\rangle = \sum_{m=1}^{d} a_m |m\rangle$$

(8)

A qutrit being a three-level (d=3) system can be expressed as

$$|\xi\rangle = a_1 |1\rangle + a_2 |2\rangle + a_3 |3\rangle$$

where $a_1, a_2, a_3 \neq 0$ and satisfy the normalization condition $|a_1|^2 + |a_2|^2 + |a_3|^2$, and $|1\rangle, |2\rangle, |3\rangle$ thereby forming an orthonormal basis for a qutrit. Representation in a three dimensional Hilbert space allows an n-qutrit system to have $3^n$ different states simultaneously.

Considering the space of states spanned by the basis vectors $|i \ j \ k---n\rangle = |i\rangle \otimes |j\rangle \otimes |k\rangle \otimes---\otimes|n\rangle$, where i j, k------n are 1, 2, or 3, an arbitrary initial state $|\psi\rangle$ of n qutrits can be expressed as

$$|\psi\rangle = \sum_{i=1}^{3}\sum_{j=1}^{3}\sum_{k=1}^{3}----\sum_{n=1}^{3} a_{i \ j \ k-n} |i \quad j \quad k---n\rangle \tag{9}$$

where normalization coefficient $a_{ijk-n}$ satisfy the condition $\sum_{ijk---n}\left|a_{ijk---n}\right|^2 = 1$. We aim to obtain an entanglement measure through $P_{\max}(\psi)$. For the same we consider a general product state of n qutrits:

$$|e\rangle = |e_1\rangle \otimes |e_2\rangle \otimes ----\otimes|e_n\rangle. \tag{10}$$

A single qutrit ignoring global phases can be written as

$$|e_k\rangle = e^{i\chi_k}\sin\theta_k\cos\gamma_k|1\rangle_k + e^{i\chi_k'}\sin\theta_k\sin\gamma_k|2\rangle + \cos\theta_k|3\rangle_k \tag{11}$$

where $0 \le \chi_k, \chi_k' \le 2\pi$ $0 \le \theta_k, \gamma_k \le \pi/2$ and $0 \le \chi_k, \chi_k' \le 2\pi$.

The product state $|e\rangle$ can thus be written as

$$|e\rangle = |e_1\rangle \otimes ----\otimes|e_n\rangle$$
$$= e^{i\chi_1}\sin\theta_1\cos\gamma_1 ---e^{i\chi_n}\sin\theta_n\cos\gamma_n|1----1\rangle + e^{i\chi_1}\sin\theta_1\cos\gamma_1---e^{i\chi_n'}\sin\theta_n\sin\gamma_n|1---2\rangle + --$$
$$+----+\cos\theta_1---\cos\theta_n|3----3\rangle \tag{12}$$

The overlap between the initial state $|\psi\rangle$ and product state $|e\rangle$, given by $\langle e|\psi\rangle$, is utilized to obtain Groverian measure of entanglement. This incorporates the maximization of the function

$$P\left(\theta_1--\theta_n, \gamma_1--\gamma_n, \chi_1--\chi_n, \chi_1'--\chi_n', \psi\right) = \left|\langle e|\psi\rangle\right|^2 \tag{13}$$

with respect to variables $\theta_k, \gamma_k, \chi_k$ and $\chi_k'$, k = 1-----n.

The maximum success probability thus becomes

$$P_{\max}(\psi) = \max_{\substack{\theta_1---\theta_n, \gamma_1----\gamma_n, \\ \chi_1---\chi_n, \chi_1'----\chi_n'}} P\left(\theta_1---\theta_n, \gamma_1---\gamma_n, \chi_1---\chi_n, \chi_1'---\chi_n', \psi\right) \tag{14}$$

upto a correction term of order $1/\sqrt{N}$ and the range of maximization is $0 \le \theta_k \gamma_k \le 2$ and $0 \le \chi_k \chi_k' \le 2\pi$. $P_{\max}(\psi)$ can be obtained by maximizing P in eq.(13) with respect to variations in $\theta_k, \gamma_k, \chi_k$, and $\chi_k'$ and equating them to zero i.e.,

$$\frac{\partial P}{\partial \theta_k} = \frac{\partial P}{\partial \gamma_k} = \frac{\partial P}{\partial \chi_k} = \frac{\partial P}{\partial \chi_k'} = 0, \tag{15}$$

for k = 1,2,-----n.

For simplicity, the phase factors involved in eq.(14) can be considered as constant terms. This facilitates the derivation of $P_{\max}(\psi)$.

## A two-qutrit system

The formulation outlined above is applied here to a two qutrit system. A pure two qutrit quantum state $|\psi\rangle$ is written as

$$|\psi\rangle = \sum_{i=1}^{3}\sum_{j=1}^{3} a_{ij}|ij\rangle \tag{16}$$

Hence an overlap of $|\psi\rangle$ with a general product state of two single qutrits, as given by eq.(11) is

$$P(\theta_1,\theta_2,\gamma_1,\gamma_2,\psi) = |\langle e|\psi\rangle|^2 =$$

$[a_{33}\cos\theta_1\cos\theta_2 + (a_{31}e^{i\chi_2}\cos\gamma_2 + a_{32}e^{i\chi_2'}\sin\gamma_2)\cos\theta_1\sin\theta_2 + (a_{13}e^{i\chi_1}\cos\gamma_1 + a_{23}e^{i\chi_1'}\sin\gamma_1)$

$\sin\theta_1\cos\theta_2 +$

$(a_{11}e^{i(\chi_1+\chi_2)}\cos\gamma_1\cos\gamma_2 + a_{12}e^{i(\chi_1+\chi_2')}\cos\gamma_1\sin\gamma_2 + a_{21}e^{i(\chi_1'+\chi_2)}\sin\gamma_1\cos\gamma_2 + a_{22}e^{i(\chi_1'+\chi_2')}\sin\gamma_1\sin\gamma_2)$

$\sin\theta_1\sin\theta_2]^2 \tag{17}$

This can be simplified by assuming $\gamma_1 + \gamma_2 = \gamma_x$ and $\gamma_1 - \gamma_2 = \gamma_y$.

$$P(\theta_1,\theta_2,\gamma_1,\gamma_2,\gamma_x,\gamma_y,\psi) =$$

$[a_{33}\cos\theta_1\cos\theta_2 + (a_{31}e^{i\chi_2}\cos\gamma_2 + a_{32}e^{i\chi_2'}\sin\gamma_2)\cos\theta_1\sin\theta_2 + (a_{13}e^{i\chi_1}\cos\gamma_1 + a_{23}e^{i\chi_1'}\sin\gamma_1)$

$\sin\theta_1\cos\theta_2 +$

$\left(\dfrac{a_{11}e^{i(\chi_1+\chi_2)} - a_{22}e^{i(\chi_1'+\chi_2')}}{2}\cos\gamma_x + \dfrac{a_{11}e^{i(\chi_1+\chi_2)} + a_{22}e^{i(\chi_1'+\chi_2')}}{2}\cos\gamma_y + \dfrac{a_{21}e^{i(\chi_1'+\chi_2)} + a_{12}e^{i(\chi_1+\chi_2')}}{2}\sin\gamma_x +\right.$

$\left.\dfrac{a_{21}e^{i(\chi_1'+\chi_2)} - a_{12}e^{i(\chi_1+\chi_2')}}{2}\sin\gamma_y\right)\sin\theta_1\sin\theta_2\Bigg]^2$

$\tag{18}$

Then by solving $\dfrac{\partial P}{\partial\gamma_1} = \dfrac{\partial P}{\partial\gamma_2} = \dfrac{\partial P}{\partial\gamma_x} = \dfrac{\partial P}{\partial\gamma_y} = 0$, maximum is found for $\gamma_1,\gamma_2,\gamma_x$ and $\gamma_y$. The values obtained are substituting in eq.(18).

Now considering $\theta_1 + \theta_2 = \theta_p$, and $\theta_1 - \theta_2 = \theta_m$, an then solving $\dfrac{\partial P}{\partial\theta_p} = \dfrac{\partial P}{\partial\theta_m} = 0$, $P_{\max}(\psi)$ is finally obtained in the following form:

$$P_{\max}(\psi) = \frac{1}{4}\Bigg[\Bigg[\bigg\{a_{33} - \frac{1}{2}\bigg(\sqrt{\left(a_{11}e^{i(\chi_1+\chi_2)} - a_{22}e^{i(\chi_1'+\chi_2')}\right)^2 + \left(a_{21}e^{i(\chi_1'+\chi_2)} + a_{12}e^{i(\chi_1+\chi_2')}\right)^2} +$$

$$\sqrt{\left(\left(a_{11}e^{i\left(\chi_1+\chi_2\right)}+a_{22}e^{i\left(\chi_1'+\chi_2'\right)}\right)^2+\left(a_{21}e^{i\left(\chi_1'+\chi_2\right)}-a_{12}e^{i\left(\chi_1+\chi_2'\right)}\right)^2\right)}\right\}^2+\left\{\sqrt{a_{13}^2e^{2i\chi_1}+a_{23}^2e^{2i\chi_1'}}+\sqrt{a_{31}^2e^{2i\chi_2}+a_{32}^2e^{2i\chi_2'}}\right\}^2\right]^{\frac{1}{2}}$$

$$+\left[\left\{a_{33}+\frac{1}{2}\left(\sqrt{\left(a_{11}e^{i\left(\chi_1+\chi_2\right)}-a_{22}e^{i\left(\chi_1'+\chi_2'\right)}\right)^2+\left(a_{21}e^{i\left(\chi_1'+\chi_2\right)}+a_{12}e^{i\left(\chi_1+\chi_2'\right)}\right)^2}+\right.\right.\right.$$

$$\left.\left.\left.\sqrt{\left(a_{11}e^{i\left(\chi_1+\chi_2\right)}+a_{22}e^{i\left(\chi_1'+\chi_2'\right)}\right)^2+\left(a_{21}e^{i\left(\chi_1'+\chi_2\right)}-a_{12}e^{i\left(\chi_1+\chi_2'\right)}\right)^2}\right)\right\}^2+\left\{\sqrt{a_{13}^2e^{2i\chi_1}+a_{23}^2e^{2i\chi_1'}}-\sqrt{a_{31}^2e^{2i\chi_2}+a_{32}^2e^{2i\chi_2'}}\right\}^2\right]^{\frac{1}{2}}\right]^2$$

$$(19)$$

Hence $P_{\max}\left(\psi\right)$ and eventually $G\left(\psi\right)$ can be calculated for any two qutrit system by substituting the value of normalized coefficients $a_{ij}$ where i, j $\in\{1, 2, 3\}$ and phase angles $\chi_1, \chi_1', \chi_2$ and $\chi_2'$.

Considering only those states $|\psi\rangle$ which have real amplitudes, the expression for $P_{\max}\left(\psi\right)$ gets simplified to

$$P_{\max}\left(\psi\right)=\frac{1}{4}\left[\left[\left\{a_{33}-\frac{1}{2}\left(\sqrt{\left(a_{11}-a_{22}\right)^2+\left(a_{21}+a_{12}\right)^2}+\sqrt{\left(a_{11}+a_{22}\right)^2+\left(a_{21}-a_{12}\right)^2}\right)\right\}^2+\right.\right.$$

$$\left.\left\{\sqrt{a_{13}^2+a_{23}^2}+\sqrt{a_{31}^2+a_{32}^2}\right\}^2\right]^{\frac{1}{2}}+\left[\left\{a_{33}+\frac{1}{2}\left(\sqrt{\left(a_{11}-a_{22}\right)^2+\left(a_{21}+a_{12}\right)^2}+\sqrt{\left(a_{11}+a_{22}\right)^2+\left(a_{21}-a_{12}\right)^2}\right)\right\}^2$$

$$\left.+\left\{\sqrt{a_{13}^2+a_{23}^2}-\sqrt{a_{31}^2+a_{32}^2}\right\}^2\right]^{\frac{1}{2}}\right]^2$$

$$(20)$$

The product state $|e\rangle$ for such states has real amplitudes, i.e., all the $\chi_k$ and $\chi_k'$ are 0 or $\pi$, leading to $\exp(i\,\chi_k)=\exp(i\,\chi_k')=\pm1$. This can be removed by doubling the range of $\theta_k$ to $-\pi/2\le\theta_k\le\pi/2$, thus $\sin\theta_k$ can be both positive and negative for the same value of $\cos\theta_k$.

Eventually, Groverian entanglement of a state $|\psi\rangle$ can be calculated by

$$G\left(\psi\right)=\sqrt{1-P_{\max}\left(\psi\right)}$$

This measure can quantify the entanglement present in various two qutrit states. Also it ca very well categorize an entangled and a product state. As for product states $P_{\max}\left(\psi\right)=1$, whereas it is never so for any entangled state. The importance of an entanglement measure lies in the fact that variations in the amount of entanglement in quantum states affects quantum computation and information processing.

For a product state,

$$G(\psi) = \sqrt{1 - P_{max}(\psi)}$$

It is easy to find out $P_{max}(\psi) = 1$, which leads to zero entanglement. This is in correspondence with one of the criteria of an entanglement measure which states $G(\psi)$ should vanish for product states.

$P_{max}(\psi)$ and finally Groverian entanglement is calculated below for certain two-qutrit systems. For a maximally entangled state of the type

$$|\psi\rangle = \frac{1}{\sqrt{3}}\left(|11\rangle + |22\rangle + |33\rangle\right)$$

$P_{max}(\psi)$=0.6666, and $G(\psi)$=0.8165.

For another extremally entangled state,

$$|\psi\rangle = \frac{1}{\sqrt{2}}\left(|11\rangle + |22\rangle\right)$$

$P_{max}(\psi)$=0.5 and thus $G(\psi)$=0.7071. For the same state the value of entanglement as reported in [19] is 0.63093.

## Conclusion

A quantum system belonging to a 3-D Hilbert space is an optimal one [25] in view of quantum computation and communication. Thus quantification of entanglement for such system facilitates its use in various information processing tasks. The expression for $P_{max}(\psi)$ as obtained above easily gives the value of $G(\psi)$ just sby inserting the value of normalized coefficients in the expression. The value of $G(\psi)$ as calculated for certain states are on expected lines. The expression for $P_{max}(\psi)$, being conceptually based on success probability of Grover's unsorted database search, indicates that entanglement is generated, rises to a maximum, and then finally vanishes during the processing of Grover's algorithm. Same can be exemplified with a two qutrit product state as the initial state of search algorithm. Unlike a two qubit product state as the starting state for the search algorithm to proceed, a two qutrit product state has maximum probability of reaching the desired state after two iterations. Entanglement is zero for the initial state, then rises to a certain value after first iteration and finally decays after second iteration. The starting state $|\psi\rangle$, a product of two qutrits in uniform superposition is

$$|\psi\rangle = \frac{1}{3}\left(|11\rangle + |12\rangle + |13\rangle + |21\rangle + |22\rangle + |23\rangle + |31\rangle + |32\rangle + |33\rangle\right)$$

Operators $P_W = 1 - 2|W\rangle\langle W|$, $|W\rangle$ being the desired state and then $P_\psi = 2|\psi\rangle\langle\psi| - 1$ (the combination of the two constitutes one Grover iteration) are applied to $|\psi\rangle$. Two Grover iterations complete the search process. The analytical expression for $P_{max}(\psi)$ also verifies that if the initial state of Grover's search algorithm is an entangled one then the performance of search algorithm is deteriorated. Entanglement being a vital part of quantum computation and communication can be exploited to the fullest only if its value is known for any quantum state

under consideration. The expression for $P_{max}(\psi)$ derived above solves the purpose well for a two qutrit system.

Figure1. shows the evolution of entanglement as the Grover's search algorithm proceeds. The solid line shows the quantification of entanglement as calculated from the expression for $P_{max}(\psi)$ whereas the dotted line quantifies entanglement on the basis of entropy of entanglement. This quantification is done by the Von Neumann entropy with

$$S_{\psi} = -Tr\left[\left(\rho_{\psi}\right)_i \log_3 \left(\rho_{\psi}\right)_i\right]$$

where i = 1 or 2 for a two qutrit system, and $\left(\rho_{\psi}\right)_i$ (i = 1 or 2) is the reduced density matrix with particle 2 or 1, respectively, traced out. Degree of entanglment has been calculated for all the intermediate states of two qutrit search algorithm generated after applying $P_W$ and $P_{\psi}$ operators to the initial state $|\psi\rangle$ within two grover iterations.

Keeping in mind the advantages of a three level system over a two level sytem, researchers have been engaged in the physical realization of qutrit systems. Quantum information processing in a qutrit system can be implemented by NMR spectroscopy [28]. Being a three level system, a qutrit can be realized in a nucleus with three energy levels. Deuterium$\left(^2H\right)$ which has nuclear spin-1 and significant quadrupole moment, is normally used in NMR spectroscopy. One qutrit quantum computer has been realized in trapped ions[29]. An axial magnetic field gradient is applied across an ion chain that allows splitting into three hyperfine Zeeman energy levels of each ion. In [30] $^{171}Yb^+$ ion with F=1 (F is for the sum of the orbital, electron-spin and nuclear-spin momenta) hyperfine Zeeman levels has been utilized as qutrit. N ions in a linear ion trap in the presence of a magnetic field gradient lead to unequally spaced hyperfine Zeeman levels which serves as qutrit. In addition, entangled qutrits for quantum communication have been experimentally realized [31] by passing an entangled photon pair through a multi-armed interferometer. The number of arms refers to the number of dimensions of the quantum system. Recently [32], arbitrary qutrit states were realized by working upon polarization state of biphoton field. This field consists of pairs of correlated photons, having equal frequencies and propagate along the same direction, obtained with the help of spontaneous parametric down conversion.

### Acknowledgement

Authors are thankful to Dr. M. D. Tiwari for his keen interest and support. Arti Chamoli is thankful is thankful to IIIT, Allahabad for financial support.

### Refrences:
[1]  C. M. Caves and G. J. Milburn, Opt. Comm. **179**, 439 (2000).
[2]  D. Bruss and C. Macchiavello, Phys. Rev. Lett. **88**, 127901 (2002).
[3]  R. T. Thew, K. Nemoto, A. G. White, and W. J. Munro, quant-ph/0201052.
[4]  J. Bouda and V. Buzek, Journal of Phys. A Math. Gen. **34**, 4301 (2001).
[5]  M. Dusek, quant-ph/0107119.
[6]  N. J. Cerf, S. Massar, and S. Pironio, quant-ph/0107031.
[7]  S. D. Bartlett, H. de Guise, and B. C. Sanders, quant-ph/0109066.
[8]  C. Brukner, M. Zukowski, and A. Zeilinger, quant-ph/0205080.
[9]  V. M. Kendon, K. Zyczkowski, and W. J. Munro, quant-ph/0203037.

[10] P. Rungta, W. J. Munro, K. Nemoto, P. Deuar, G. J. Milburn, and C. M. Caves, quant-ph/0001075.

[11] V. Vedral, M. B. Plenio, M. A. Rippin, and P. L. Knight, Phys. Rev. Lett. **78**, 2275 (1997).

[12] S. Hill and W. K. Wootters, Phys. Rev. Lett. **78**, 5022 (1998).

[13] W. K. Wootters, Phys. Rev. Lett. **80**, 2245 (1998).

[14] V. Vedral and M. B. Plenio, Phys. Rev. A **57**, 1619 (1998).

[15] V. Coffman, J. Kundu, and W. K. Wootters, Phys. Rev. A **61**, 052306 (2000).

[16] A. Wong and N. Christensen, Phys. Rev. A **63**, 044301 (2001).

[17] G. Vidal and R. F. Werner, Phys. Rev. A **65**, 032314 (2002).

[18] W. Dur, G. Vidal, and J. I. Cirac, Phys. Rev. A **62**, 062314 (2000).

[19] F. Pan, Guoying Lu, and J. P. Drayer, quant-ph/0510178.

[20] F. Pan, D Liu, G. Y. Lu, and J. P. Drayer, Int. J. Theor. Phys. **43**, 1241 (2004).

[21] A. Chamoli and C. M. Bhandari, Phys. Lett. A **346,** 17 (2005).

[22] N. J. Cerf, M. Bourennane, A. Karlsson, and N. Gissin, Phys. Rev. Lett. **88**, 127902 (2002).

[23] T. Durt, N. J. Cerf, N. Gissin, and M. Zukowski, Phys. Rev. A **67**, 012311 (2003).

[24] R. W. Spekkens and T. Rudolph Phys. Rev. A **65**, 012310 (2002).

[25] A. D. Greentree, S. G. Schirmer, F Green, L. C. L. Hollenberg, A. R. Hamilton and R. G. Clark, Phys. Rev. Lett. **92**, 097901 (2004).

[26] O. Biham, M. A. Nielsen, and T. Osborne, Phys. Rev. A. **65**, 062312 (2002).

[27] Y. Shimoni, D. Shapira, and O. Biham, Phys. Rev. A. **69**, 062303 (2004).

[28] R. Das, A. Mitra, V. Kumar S, and A. Kumar, quant-ph/0307240.

[29] A. B. Klimov, R. Guzmn, J. C. Retamal, and C. Saavedra, Phys. Rev. A **67**, 062313 (2003).

[30] D. Mc Hugh and J. Twamley, New J. Phys. **7**, 174 (2005).

[31] R. T. Thew, A. Acin, H. Zbinden, and N. Gisin, Quantum Information and Computation, **Vol.4**, No.2, 93 (2004).

[32] Y. I. Bogdanov, M. Chekhova, S. Kulik, G. Maslennikov, C. H. Oh, M. K. Tey, and A. A. Zhukhov, Phys. Rev. Lett., **93**, 230503 (2004).

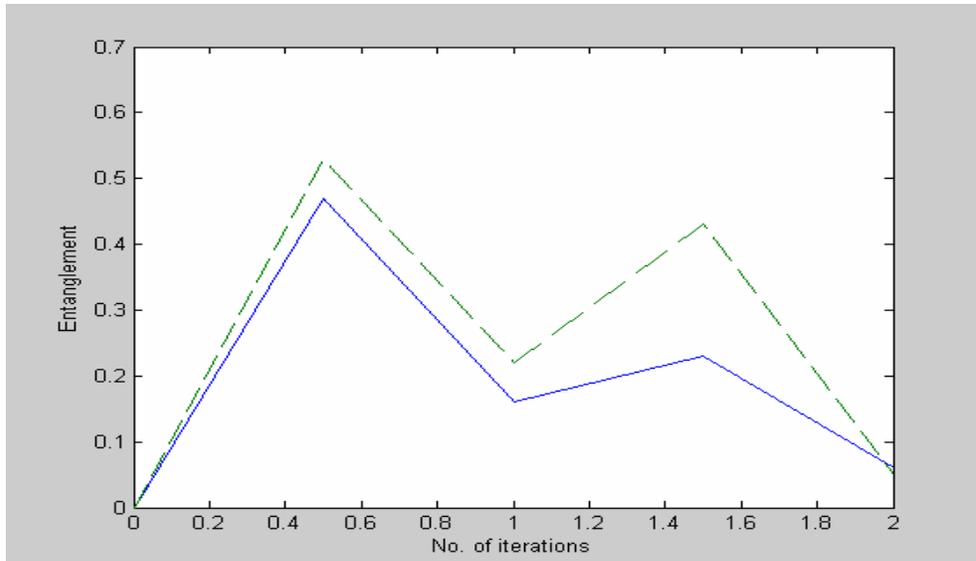

FIG1. Evolution of entanglement for a two qutrit system with application of Grover operators. Solid line indicates Groverian measure; dotted line indicates entanglement measure as entropy of entanglement.